\documentclass[11pt]{article}
\usepackage{amsfonts}
\usepackage{amsmath}
\usepackage{amssymb}
\usepackage{amsthm}
\usepackage{graphicx}
\usepackage{indentfirst}

\newenvironment{Proof}
{\par\noindent{\bf Proof.}}
{\hfill$\scriptstyle\blacksquare$\medskip}

\newtheorem{lemma}{Lemma}
\newtheorem*{Theorem}{Theorem}
\newtheorem{Def}{Definition}

\usepackage{tikz}
\usetikzlibrary{positioning}
\usetikzlibrary{backgrounds}
\usetikzlibrary{fit}
\usetikzlibrary{arrows}

\newcommand{\defemph}[1]{\textbf{#1}}

\begin{document}

\author{Raskin M.A.\footnote{Aarhus University, CS Dept. email: \texttt{raskin@cs.au.dk}, \texttt{raskin@mccme.ru}. The work was partially supported by RFBR grant 16-01-00362. The author acknowledges support from the Danish National Research Foundation
and The National Science Foundation of China (under the grant 61361136003) for
the Sino-Danish Center for the Theory of Interactive Computation and from the Center for
Research in Foundations of Electronic Markets (CFEM), supported by the Danish
Strategic Research Council.} \and Nikitenkov N.S.\footnote{Moscow State University, Mech. and Math. Dept.}}

\title{ Paradoxical examples of social networks games with product choice}

\maketitle

\tikzstyle{spirit}=[scale=1, circle, draw=black, fill=gray!20, very thick, inner sep=1pt, minimum size=8mm]
\tikzstyle{human}=[circle, draw=black, fill=blue!20, very thick, inner sep=1pt, minimum size=8mm] 
\tikzstyle{cascade}=[scale=1.5, rectangle, draw=black, fill=blue!20, very thick,inner sep=10pt, rounded corners] 
\tikzstyle{pre}=[<-, very thick]
\tikzstyle{post}=[->, very thick]
\tikzstyle{slight}=[->, shorten >=3pt,>=stealth', shorten <=3pt,>=stealth', very thick, dashed, draw=black!70]
\tikzstyle{pslight}=[<-, shorten >=3pt,>=stealth', shorten <=3pt,>=stealth', very thick, dashed, draw=black!70]
\tikzstyle{every fit}=[scale=1, fill=black!10,rounded corners,minimum size=20mm]
\tikzstyle{invisible}=[]


\begin{abstract}

Paradox of choice occurs when permitting new strategies to some
players yields lower payoffs for all players in the new equilibrium
via a sequence of individually rational actions.

We consider social network games. 
In these games the payoff of each player increases when other
players choose the same strategy. 

The definition of games on social networks was introduced 
by K. Apt and S. Simon.  In an article written jointly with E. Markakis, 
they considered four types of paradox of choice in such games
and gave examples of three of them. 
The existence of paradoxical networks of the fourth type
was proven only in a weakened form. The existence of so-called
«vulnerable networks» in the strong sense remained an open question. 

In the present paper we solve this open question
by introducing a construction, called a «cascade», 
and use it to provide uniform examples for all four definitions of paradoxical 
networks. 
\end{abstract}

\section{Introduction}

One of the topics of interest in game theory is the paradox of choice. In formal game theory this
name applies to a situation where permitting new strategies to one or more players leads to some
rational selfish decisions and eventually worsens situation (yields lower payouts) for all players
in the new Nash equilibrium.

Perhaps the most famous paradox of that type is Braess's paradox \cite{braess}. 
In this game many players try 
to get from one node of a graph to another in a graph with edges having different bandwidths.
It turns out that adding additional edges can lead to increased transit time for all players.

In the present paper we consider another subclass of games on graphs, the so-called 
social network games. In these games the payoff of each player increases when other
players choose the same strategy. We will define the game rigorously later.

The definition of games on social networks was introduced in the article \cite{book1} 
by K. Apt and S. Simon. They have obtained a number of theoretical results 
(in particular, it was shown that the problem of finding a Nash equilibrium in a social 
network game in the general case is NP-complete). In a later article \cite{book2}, 
written jointly with E. Markakis, the authors considered four hypothetical types of
paradox of choice in games on social networks and gave examples of three of them. 
The existence of paradoxical networks of the fourth type, the so-called vulnerable 
networks, was proven only in a weakened form. The question of the existence of
vulnerable networks in the strong sense was left open. A vulnerable network always 
provides an example of paradox of choice as defined above.

In the present article we solve the open problem of Apt, Markakis and Simon.
We introduce a construction, called a «cascade», 
and use it to provide uniform examples for all four definitions of paradoxical 
networks. 
In addition, we use cascades to construct a number
of other examples, which also can be considered paradoxical, though in 
a slightly different sense.

\section{Basic definitions}

\subsection{Social network}

Let $V = \{1,\dots, N\}$ be a finite set of nodes. We consider a directed graph 
with weighted edges $G = (V, E, \omega)$ with non-negative edge weights: $\omega_{ij} \geqslant 0$. 
We assume that the weight of a missing edge is zero.
Let $\Pi$ be a finite set of possible individual strategies or \defemph{products}. 
Let $P: V \to 2^\Pi$ be a function describing \defemph{ available} products 
(i.e. allowed strategies) for every player.
Let $\theta$ be a \defemph{threshold function}. For each player $i \in V$ and
each product $t \in P(i)$ the value $\theta(i, t) > 0$ is called \defemph{use price}.
\begin{Def}
A \defemph{social network} (or simply a network) is a set $S = (G, \Pi, P, \theta)$.
\end{Def}

\begin{Def}
A network $S'$ is called  an \defemph{expansion} of the network $S$ if it is obtained 
by allowing a single additional product for a single node. 
The network $S$ is called a \defemph{contraction} of the network $S'$.
\end{Def}

\subsection{The game on a social network}
Now we consider a strategic game with each node 
of the social network $S$ being an independent player. 
All players simultaneously make a choice between use of one of the 
available products $t_i \in P(i)$ or \defemph{ refusal} 
to use any product (we will call this choice $t_0$). 
Thus, for each player $i \in V$ we have defined a set of 
strategies $S_i = P(i) \cup \{t_0\}$.

We will use the notation $G(S)$ to denote the strategic game based on $S$. 
We will call a strategy profile of all players $s \in \prod_{i \in V} S_i$ 
a \defemph{ position} of the game $G(S)$. 
In a position $s$, let $s_i$ denote the individual strategy for player $i$ in $s$, and 
let $N(t)$, where $t \in \Pi$, denote the set of all players who have selected the product $t$.

Each player will pay the price of the selected product and will
receive utility proportional to the incoming link weight whenever an
acquaintance has chosen the same product.
That is, we define the payoff function $p_i$ in the following way:
\[
p_i(s) = \begin{cases}
0, & \text{if $s_i = t_0$} \\
(\sum_{j \in N(t)} \omega_{ij}) - \theta(i, t), & \text{if $s_i = t \in P(i)$}
\end{cases} 
\]

\begin{Def}
Let two positions, $s$ and $s'$ differ only by the individual strategy for player $i$:
$\forall j \not= i: s_j = s _j$. Let $p_i(s) < p_i(s')$. 
Then we say that $s'$ is obtained by an \defemph{ individual improvement} of $s$. 

If a position $s$ allows no individual improvements, the position $s$ 
is a \defemph{ Nash equilibrium} for the game $G(S)$.

An \defemph{individual improvement path} is a sequence of positions, 
in which each next position is an individual improvement on the previous one.

\end{Def}

\section{Paradoxes in games on social networks}
Given two states $s$ and $s'$ in a game $G(S)$ we write 
that $s > s'$ if $\forall i \in V: p_i(s) > p_i(s')$.

\begin{Def}
A network $S$ is called \defemph{ vulnerable}, if 
for some Nash equilibrium state $s$ in $G(S)$ there exists 
an expansion network $S'$ such that each individual improvement path 
in $G(S')$ starting at the state $s$ leads to a state $s'<s$, 
which is a Nash 
equilibrium in both $G(S)$ and $G(S')$.
\end{Def}

\begin{Def}
A network $S$ is called \defemph{ fragile}, if for some
Nash equilibrium state $s$ in $G(S)$ there exists an expansion 
network $S'$ such that each individual improvement path in $G(S')$
starting at the state $s$ is infinite.
\end{Def}

\begin{Def}
A network $S$ is called \defemph{ ineffective}, if for some Nash equilibrium 
state $s$ in $G(S)$ there exists a contraction $S'$ such that each 
individual improvement path in $G(S')$ 
starting at the state $s$ leads to a state $s' > s$ which is a Nash 
equilibrium in both $G(S)$ and $G(S')$.
\end{Def}

\begin{Def}
A network $S$ is called \defemph{ unsafe}, if 
for some Nash equilibrium state $s$ in $G(S)$ there exists
a contraction $S'$ such that any individual improvement path in $G(S')$
starting at the state $s$ is infinite.
\end{Def}

\begin{Theorem}
There are networks of all four types described.
\end{Theorem}

This theorem has been partially proved in \cite{book2}. 
The authors give isolated examples of fragile, inefficient and unsafe networks, 
and prove the existence of vulnerable networks in a weaker sense, 
namely for social games in which players are not allowed 
to refuse to use any product.

Before giving the proof of the theorem, we will consider 
a network design that we call a \defemph{cascade}. 
We will use it to construct an example of a vulnerable network 
in the original sense and also a uniform series of examples 
for three other paradox types.

\section{A cascade}

\defemph{A cascade} is a social network with three products.
It can include an arbitrarily large number of players and always 
has two non-trivial equilibriums states (the choice can be
affected by incoming edges). 
The transition from the first equilibrium to the second one 
leads to a severe reduction in the individual payoff of each player. 
Moreover, this payoff difference grows proportionally to the size of 
the cascade and can reach arbitrarily large values.

In the situation when a cascade is built into
a larger social network in the form of a subgraph,
changes in the payoff can also apply to some of the
vertices outside the cascade.
In a way, the cascade is 
a ``happiness machine'' that can be ``turned on'',
making all the players in the system arbitrarily happy,
and ``switched off'', depriving them of the additional utility.

We will now describe the construction of the cascade.

In a cascade there are three products denoted by A, B and C. 
In the diagrams below, we enumerate available products for each 
player directly inside the corresponding node of the graph.

\subsection{Types of players}

The cascade includes two types of players called ``humans'' and ``spirits''.

Each ``human'' player has a choice of two different products. 
These products correspond to the two different non-trivial equilibriums. 
Of course, there are six possible subtypes of human players: 
AB, AC, BA, BC, CA, CB. Every cascade includes an equal number of players 
of all six subtypes.

The number of players of one subtype will be denoted by $n$. 
The total number of human players is $6n$.

A spirit player has only one product available, which he will use in both equilibriums of the cascade. 
The cascade uses six spirits: two for each of three products.

The thresholds for all players are equal and close to zero. 
We'll use $\theta$ to denote them.

\subsection{Types of edges}
In every cascade, there are edges of three different types.

The \defemph{control} edges have the largest weight.
In a cascade, each player has at most one outgoing and one incoming control edge. 
The design of a cascade is such that each player always follows 
the incentives set by the incoming control edge whenever possible
(i.e. when the corresponding product is available).
In the diagrams the control edges will be shown by solid black arrows.
Each control edge has the same weight~$c$. 

The \defemph{inclination} edges 
determine the behavior of a player when the corresponding 
control edge incentivises the use of an unavailable product.
Each human player has exactly one incoming inclination edge
from one of six spirits. The spirits themselves do not have 
incoming inclination edges.

Since the spirits cannot change their chosen product, 
each inclination edge gives its target player a constant
preference for using some product. In the diagrams we will 
emphasise the corresponding product if it is available.

All inclination edges have the same weight $i < c$.

The remaining edges inside a cascade are called \defemph{emotional} edges.
Their individual weight is comparatively small, however, 
every player in the cascade has a lot of incoming emotional edges, 
so in total they have a significant effect. However, incoming emotional 
edges never influence a player's strategic choices. More precisely,
for each player in the cascade the following property always holds.

\medskip
\begin{Def} \emph{(emotional invariant)} 
Denote by $E(v,\alpha)$ the number of incoming emotional edges for the player $v$ 
rewarding the choice of the product $\alpha$. We will say that 
the \defemph{ emotional invariant} holds for the player $v$, 
if for each pair of products $(\alpha,\beta)$ available 
to the player $v$, $|E(v,\alpha) - E(v,\beta)| \leqslant 1$. 
Note that there are no restrictions for the value $E(v,\gamma)$ 
when the product $\gamma$ 
is not available to the player~$v$. 
\end{Def}
\medskip

Thus, all incoming emotional edges for each player are always 
balanced between all the available products (the difference is no
more than a single edge). Switching the emotional edges to reward
the choice of the unavailable products allows to reduce the eventual
payoffs.

Emotional edges will not be indicated on the diagrams. 
We will only describe their layout in the text.
All the emotional edges have the same weight, we'll call it $e$.
The following conditions will hold:
$e > 0, e < i, e < c, e < c - i$, but $en > c + i$.

\subsection{Components of the cascade}

\subsubsection{Stimuli}
Each stimulus consists of isolated an pair of identical spirits, with control edges in both directions. This coupling ensures that the spirits will never refuse to use the product available to them.

\medskip
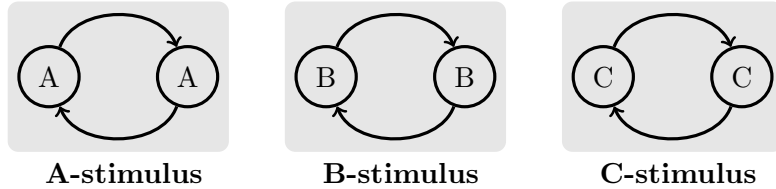
\begin{center}
\begin{tikzpicture}
[bend angle=70]

\node[spirit] (a1) {A};
\node[spirit] (a2) [right=of a1] {A}
edge [pre, bend right] (a1)
edge [post, bend left] (a1);

\node[spirit] (b1) [right=of a2] {B};
\node[spirit] (b2) [right=of b1] {B}
edge [pre, bend right] (b1)
edge [post, bend left] (b1);

\node[spirit] (c1) [right=of b2] {C};
\node[spirit] (c2) [right=of c1] {C}
edge [pre, bend right] (c1)
edge [post, bend left] (c1);

\begin{scope}[on background layer]
\node [fit=(a1)(a2)] [label=below:\textbf{ A-stimulus}] {};
\node [fit=(b1)(b2)] [label=below:\textbf{ B-stimulus}] {};
\node [fit=(c1)(c2)] [label=below:\textbf{ C-stimulus}] {};
\end{scope}

\end{tikzpicture}
\end{center}

The only purpose of the spirits in the stimuli is to serve as the sources 
of inclination edges to human players. Recall that each human player has 
exactly one incloming inclination link, which determines its behavior 
in the case when following the control link is impossible.

\subsubsection{Ranks}

Each rank is a chain of human players. 
All rank members share one available product; 
each of them has another available product that alternates 
between the remaining two options. 
Rank members are connected by control edges along the rank.

All players in the line, except the first, have the inclination
towards the second (alternating) available product, and the first human
has an inclination towards the first (common) available product. 
Common product for a line of players will be called the
\defemph{ main} product for this line, and the two alternating products 
will be called \defemph{secondary}. We will refer to the ranks by
mentioning their main products. 

A cascade uses a single rank of each type with $2n$ players 
in each. The figure below shows an example of an A-rank:

\medskip
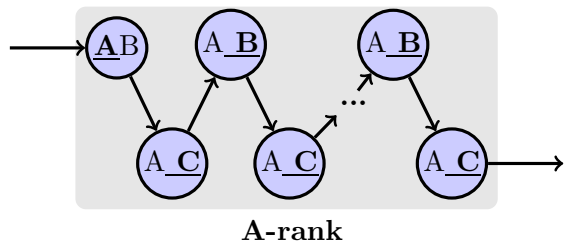
\begin{center}
\begin{tikzpicture}

	\node[human] (ab1) {\underline{\textbf{A}}B};
\node[human] (ac1) [below right =9mm and 1mm of ab1] {A\underline{\textbf{ C}}}
edge [pre] (ab1); 
\node[human] (ab2) [above right =9mm and 1mm of ac1] {A\underline{\textbf{ B}}}
edge [pre] (ac1);
\node[human] (ac2) [below right =9mm and 1mm of ab2] {A\underline{\textbf{ C}}}
edge [pre] (ab2);

\node[invisible] (i1) [above right =3mm and 0.5 mm of ac2] {\textbf{ ...}}
edge [pre] (ac2);

\node[human] (ab3) [above right =9mm and 7mm of ac2] {A\underline{\textbf{ B}}}
edge [pre] (i1);
\node[human] (ac3) [below right =9mm and 1mm of ab3] {A\underline{\textbf{ C}}}
edge [pre] (ab3);

\node[invisible] [left = of ab1] {}
edge [post] (ab1);
\node[invisible] [right = of ac3] {}
edge [pre] (ac3);

\begin{scope}[on background layer]
\node [fit=(ab1)(ac3)] [label=below:\textbf{ A-rank}] {};
\end{scope}

\end{tikzpicture}
\end{center} \pagebreak

The ranks of a cascade are sources of balanced emotional edges.
We introduce the following definition:

\begin{Def}
We say that rank \defemph{ has emotional influence} on player $v$, if each player in the rank is connected to $v$ with one 
outgoing emotional edge. 
The rank can have an emotional influence on several players, 
both inside and outside of the cascade.
\end{Def}

Each player in the rank, in turn, is under external emotional 
influence (subject to emotional invariant). In addition, the 
first player in the rank has an incoming control edge from some
external player and the last player has an outgoing control edge 
to some external player (or players). The players inside a rank 
have no other external connections.

Consider the possible states of a rank in the assumption that 
the emotional invariant set forth above is not violated (in other 
words, incoming emotional edges will not affect the strategic 
choices of players).

\begin{lemma}
\emph{(on rank states)}
Assume that the emotional invariant holds for every player in 
a rank.

Also assume that initially either everyone uses the main product
or everyone uses the secondary product.

Then the choice of the first player in the rank between the main 
and the secondary product becomes the choice of the entire row 
as a result of the only possible chain of individual 
improvements.
\end{lemma}
\begin{Proof}
First, note that if the first player in the line uses the main 
product (at least because of its inclination edge), under the 
influence of the control edges all the rank will also be forced 
to use the main product (despite their inclinations).

Now consider a chain of improvements if the first player 
will change its strategy in favor of a secondary product. 
Then the next player will not be able to submit to the control 
edge and so will also change their choice in favor of 
the secondary product under the effect of incoming 
inclination link. This in turn will make it impossible for 
the third player to follow the incoming control edge and so on.
As a result all the rank will consistently change their choice 
in favor of secondary products. Obviously, this is the only 
possible chain of improvements in the described case.

The converse is also true: if you remove the external influence 
on the first player, then the only possible result of a chain 
of improvements is for all the rank to return to the main product.
\end{Proof}

We will call the \defemph{ first state of a rank} 
the of the main product by all the players, and the 
\defemph{ second state of a rank}, respectively, 
the use of secondary products.

Now we prove the main property of a rank:

\begin{lemma}
\emph{(on maintaining emotional invariant)}
Let a rank with the main product $\alpha$ and the secondary 
products $\beta$ and $\gamma$ have an emotional influence
on an external player $v$. Then in both equilibrium states of 
the rank and at any step during the chain of improvements 
leading from one to another, 
$|E(v,\beta) - E(v,\gamma)| \leqslant 1$.
\end{lemma}
\begin{Proof}
In the first state of the rank none of its constituent players 
uses the products $\beta$ and $\gamma$ and the specified property 
is trivially true. 
In the second state of the rank exactly half of the players 
use the product $\beta$, and the remaining half use
the product $\gamma$, so $E(v,\beta) = E(v,\gamma)$.
During the transition from the first state to the second one,
players in line sequentially switch from $\alpha$ 
to $\beta$, then from $\alpha$ to $\gamma$, and hence 
the magnitude of $E(v,\beta)$ and $E(v,\gamma)$ in the queue 
increases by one. Similarly, during the transition from the 
second state of the rank to the first one, 
the players sequentially switch from $\beta$ and $\gamma$ 
to $\alpha$, hence the value of $E(v,\beta)$ and $E(v,\gamma)$ 
in the queue decreases by one. 
In both cases $|E(v,\beta) - E(v,\gamma)|$ stays
less than or equal to 1 all the time.
\end{Proof}

Note that if the player $v$ can use only the products 
$\beta$ and $\gamma$ or only one of them 
(and no other products are available), the emotional invariant 
property means that the influence of the emotional edges from
the rank is always balanced between the available products.
In such a case in the second state the rank gives additional 
utility $ne$ to the player $v$ compared to the first state
of the rank.

\subsection{General arrangement of the cascade}
\medskip
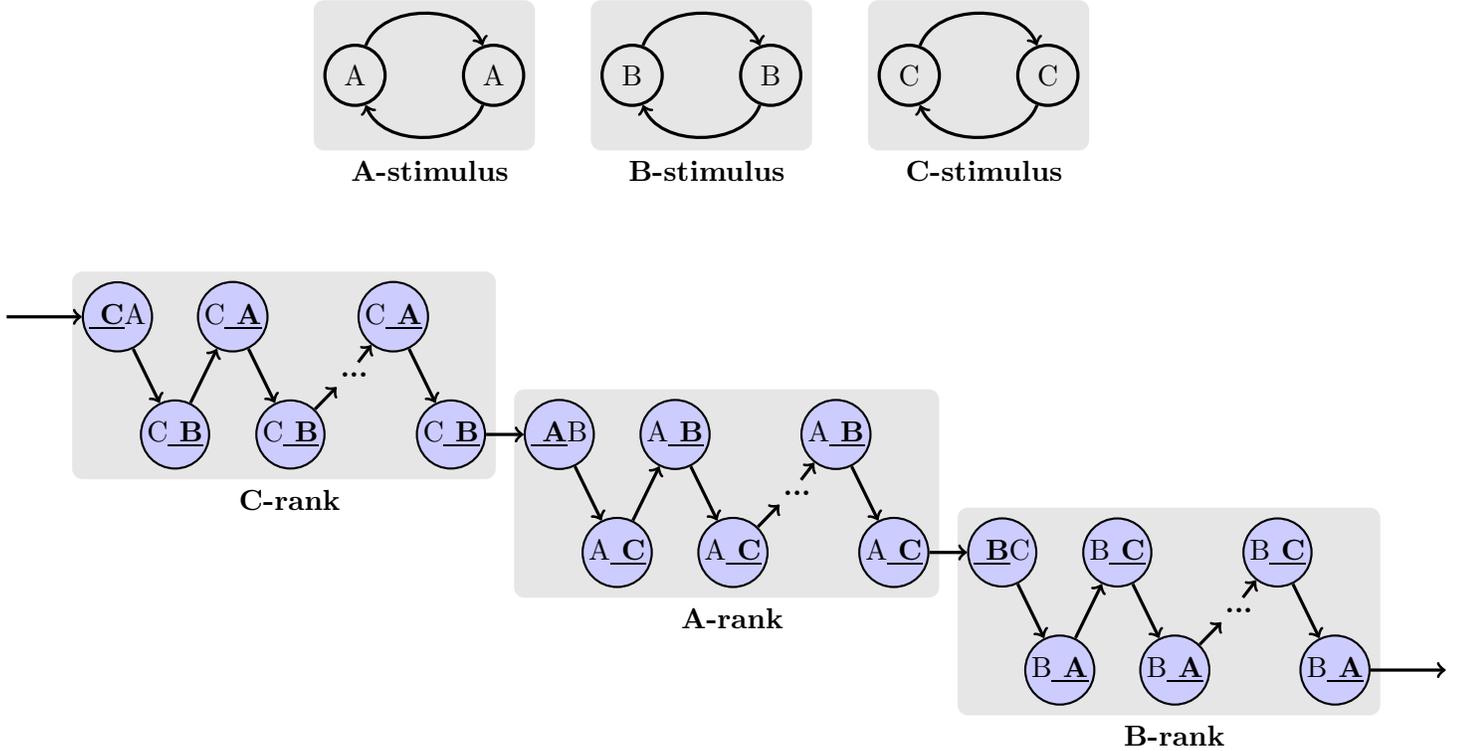
\begin{center}
\begin{tikzpicture}
[bend angle=70]
\node[spirit] (a1) {A};
\node[spirit] (a2) [right=of a1] {A}
edge [pre, bend right] (a1)
edge [post, bend left] (a1);

\node[spirit] (b1) [right=of a2] {B};
\node[spirit] (b2) [right=of b1] {B}
edge [pre, bend right] (b1)
edge [post, bend left] (b1);

\node[spirit] (c1) [right=of b2] {C};
\node[spirit] (c2) [right=of c1] {C}
edge [pre, bend right] (c1)
edge [post, bend left] (c1);

\begin{scope}[on background layer]
\node [fit=(a1)(a2)] [label=below:\textbf{ A-stimulus}] {};
\node [fit=(b1)(b2)] [label=below:\textbf{ B-stimulus}] {};
\node [fit=(c1)(c2)] [label=below:\textbf{ C-stimulus}] {};
\end{scope}

\end{tikzpicture}
\end{center}

\begin{center}
\begin{tikzpicture}
[scale=1,auto=left, align=center,
spirit/.style={scale=1, circle, draw=black, fill=gray!20, very thick},
human/.style={circle, draw=black, fill=blue!20, thick, inner sep=1pt},
invisible/.style={},
bend angle=70,
pre/.style={<-, very thick},
post/.style={->, very thick},
every fit/.style={scale=1, fill=black!10,rounded corners,minimum size=20mm}]

\node[human] (ca1) {\underline{\textbf{ C}}A};
\node[human] (cb1) [below right =9mm and 1mm of ca1] {C\underline{\textbf{ B}}}
edge [pre] (ca1);
\node[human] (ca2) [above right =9mm and 1mm of cb1] {C\underline{\textbf{ A}}}
edge [pre] (cb1);
\node[human] (cb2) [below right =9mm and 1mm of ca2] {C\underline{\textbf{ B}}}
edge [pre] (ca2);
\node[invisible] (i3) [above right =3mm and 0.5 mm of cb2] {\textbf{ ...}}
edge [pre] (cb2);
\node[human] (ca3) [above right =9mm and 7mm of cb2] {C\underline{\textbf{ A}}}
edge [pre] (i3);
\node[human] (cb3) [below right =9mm and 1mm of ca3] {C\underline{\textbf{ B}}}
edge [pre] (ca3);
\begin{scope}[on background layer]
\node [fit=(ca1)(cb3)] [label=below:\textbf{ C-rank}] {};
\end{scope}

\node[human] (ab1) [right=5mm of cb3] {\underline{\textbf{ A}}B}
edge [pre] (cb3);
\node[human] (ac1) [below right =9mm and 1mm of ab1] {A\underline{\textbf{ C}}}
edge [pre] (ab1);
\node[human] (ab2) [above right =9mm and 1mm of ac1] {A\underline{\textbf{ B}}}
edge [pre] (ac1);
\node[human] (ac2) [below right =9mm and 1mm of ab2] {A\underline{\textbf{ C}}}
edge [pre] (ab2);
\node[invisible] (i1) [above right =3mm and 0.5 mm of ac2] {\textbf{ ...}}
edge [pre] (ac2);
\node[human] (ab3) [above right =9mm and 7mm of ac2] {A\underline{\textbf{ B}}}
edge [pre] (i1);
\node[human] (ac3) [below right =9mm and 1mm of ab3] {A\underline{\textbf{ C}}}
edge [pre] (ab3);
\begin{scope}[on background layer]
\node [fit=(ab1)(ac3)] [label=below:\textbf{ A-rank}] {};
\end{scope}

\node[human] (bc1) [right=5mm of ac3] {\underline{\textbf{ B}}C}
edge [pre] (ac3);
\node[human] (ba1) [below right =9mm and 1mm of bc1] {B\underline{\textbf{ A}}}
edge [pre] (bc1);
\node[human] (bc2) [above right =9mm and 1mm of ba1] {B\underline{\textbf{ C}}}
edge [pre] (ba1);
\node[human] (ba2) [below right =9mm and 1mm of bc2] {B\underline{\textbf{ A}}}
edge [pre] (bc2);
\node[invisible] (i2) [above right =3mm and 0.5 mm of ba2] {\textbf{ ...}}
edge [pre] (ba2);
\node[human] (bc3) [above right =9mm and 7mm of ba2] {B\underline{\textbf{ C}}}
edge [pre] (i2);
\node[human] (ba3) [below right =9mm and 1mm of bc3] {B\underline{\textbf{ A}}}
edge [pre] (bc3);
\begin{scope}[on background layer]
\node [fit=(bc1)(ba3)] [label=below:\textbf{ B-rank}] {};
\end{scope}

\node[invisible] [left = of ca1] {}
edge [post] (ca1);
\node[invisible] [right = of ba3] {}
edge [pre] (ba3);

\end{tikzpicture}

\end{center}

The cascade consists of three different stimuli 
and three different ranks connected by control edges as shown in the figure.

Each rank has emotional  influence on all the players who don't have access 
to its main product.
So, the A-rank has an emotional influence on all human players types BC and CB,
and the spirits of types B and C.

\begin{lemma}
In the above construction of the cascade the state of the first rank 
is translated to the second and third ranks in the result of
the only possible chain of individual improvements.
The individual payoffs of each player in the second state are higher than 
the corresponding payoffs in the first state by at least $(ne - c - i)$.
\end{lemma}

\begin{Proof}
First note that for each player in the cascade (according to \emph{Lemma 2}) 
the emotional invariant holds. 
Therefore, if the first rank is in the first state, 
the second rank will not be able to follow incoming control edge, 
and (\emph{Lemma 1}) will also switch to the first state.
Similarly, it will be impossible for the third row to 
follow the incoming control edge, so it will also adopt the first state.

In this case, all emotional edges will push players 
to the selection of products available to them.

If the first rank moves into its second state, 
the second rank will be able to follow incoming control edge, 
and (\emph{Lemma 1}) also so it will move into 
its second state, making the third rank 
also switch to the second rank.

In this case, due to the emotional edges all the
human players receive extra 
payoffs of size $ne$, and all the spirit players get an 
additional gain of $3ne$ that obviously outweighs 
the gains from control and inclination edges.

Thus, individual payoffs of all the players in the second state 
of the cascade are greater than in the first state of the cascade 
by at least $(ne - c - i)$.
\end{Proof}

Note that since $n$ can be chosen arbitrarily high, 
the potential difference of the payoffs is unlimited.

\subsection{Properties of the cascade}
We now formulate the properties in cascade the form of a theorem:
\begin{Theorem}
\emph{(on the properties of the cascade)}
The design of the cascade has the following properties:
\begin{enumerate}
\item The cascade has one incoming control edge and one outgoing control edge.
\item If a cascade is used as a source of inclination edges to external players,
these edges incentivise selection of the same products, 
regardless of the state of the cascade.
\item There is an equilibrium state of the cascade when the 
incoming edge incentivises the use of the product A and the cascade 
is in its second state. The outgoing edge in this situation also
incentivises the use of the product A.
\item There is also an equilibrium state of the cascade when the 
incoming edge incentivises the use of any other product and the
cascade is in the first state. 
The outgoing edge in this situation 
incentivises the use of the product B.
\item If the incoming edge switches when the cascade is in one 
of the two aforementioned equilibrium states, the cascade switches 
to the other state following the only possible individual
improvements sequence.
\item In the second state the cascade gives additional emotional
payoff to all its players.
\item The individual payoffs of all the players in the second state of the cascade 
are higher than in the first state of the cascade at least by $(ne - c - i)$.
\item The emotional influence of the cascade can be extended to external players 
having access to one or two products from the set \{A,B,C\} and no other products. 
In respect of these players the emotional invariant will hold 
and they will receive additional emotional payoff in the amount of at least $ne$ 
when the cascade is in the second state, 
and no emotional payoff, when the cascade is in the first state.
\end{enumerate}
\end{Theorem}

\begin{Proof}
The cascade device and \emph{Lemmas 1, 2 and 3}.
\end{Proof}

These properties will be used when using the cascade as a subgraph
in a larger network.

\subsection{Choosing parameters}
Above we have formulated conditions on the numerical parameters of the cascade:

$\theta > 0$;

$e > 0$;

$i > e$; $i > \theta$;

$c > i + e$;

$n > \frac{c + i}{e}$;

\medskip
These conditions can be met by selecting each parameter large enough. 
In order to demonstrate the existence of the cascade in principle,
it is enough to consider $\theta = 1, e = 1, i = 3, c = 5, n = 10$.

We can also determine the minimum size of the cascade that meet the specified conditions.
\begin{lemma}
The minimum number of players in the cascade is 30.
\end{lemma}
\begin{Proof}
Combining inequality $c - i > e$ and $2i > 2e$ we get $c + i > 3e$, and 
hence the compliance is impossible when $n < 4$.

When $n = 4$ a cascade exists, for example, 
with these values: $\theta = 0.1, i = 0.3, c = 0.6, e = 0.25$.

Thus, the minimum cascade can be built with $4 \cdot 6 + 6 = 30$ players.
\end{Proof}

\section{Examples of paradoxical networks}

Now we return to the proof of the main theorem about the paradoxical networks.

\subsection{A vulnerable network}

\medskip

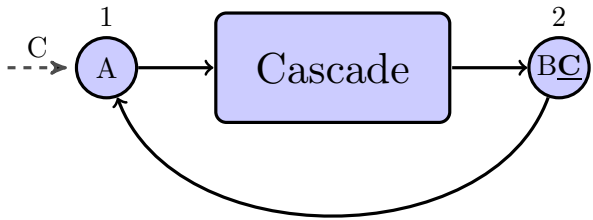
\begin{tikzpicture}
[bend angle=70]

\node[human] (a) [label=above:{1}] {A};
\node[cascade] (casc) [right=of a] {Cascade}
edge[pre] (a);
\node[human] (bc) [right=of casc] [label=above:{2}] {B\underline{\textbf{C}}}
edge[pre] (casc)
edge[post, bend left] (a);
\node[invisible] [left=of a]{}
edge [slight] node[above, swap] {C} (a);

\end{tikzpicture}

\medskip
\begin{Proof}
We add two players to a cascade
as shown in the diagram 
(solid arrows denote control relations, dotted arrow pointing to the first player 
denotes adding an additional product). 
Note that the second external player needs the inclination edge from one of the C-spirits. 
In addition, both external player must be under emotional influence from B- and A-ranks,
respectively.

Initial Nash equilibrium:
the first player uses the product A, 
the cascade under the influence of the control edge is in the second state, 
the second player can't follow the control edge,
so according to his inclination he uses the product C. 
Since the cascade is in the second state, all players (including the two 
additional players) get an additional payoff due to the emotional edges. 
Because of this extra payoff the first player will not refuse 
to use the product A.
He can't obey incoming control edge, as the product C is not available. 
Thus, the network is indeed in equilibrium.

Now allow the first player to use the product C. 
Emotional edges won't affect the choice between A and C, so he will
obey the incoming control edge and switch to the product C.
This will lead to the transition of the cascade into the first state 
and the control edge will now force the second player to switch and use 
the product B.
The control edge from the second player to the first becomes
impossible to obey for the first player. Moreover, all the incoming 
edges will reward choosing the unavailable product B. The first player
will abandon the use of any product to save on the threshold cost.
This will not lead to any further changes, so the network
will have reached a new equilibrium with strictly lower payoffs
for all players.
Note that this position is also an equilibrium in the original network.

Thus, the constructed network is really vulnerable.
\end{Proof}

Note that in \cite{book1} it was proved that a vulnerable network can't be built 
using less than three products. The presented example demonstrates that
three products are sufficient.

\subsection{A fragile network}

\medskip

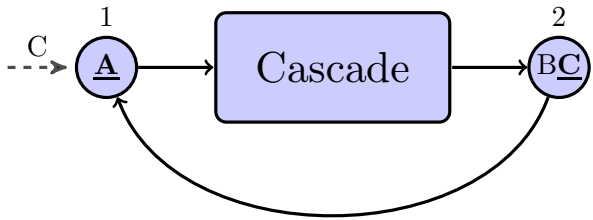
\begin{tikzpicture}
[bend angle=70]

\node[human] (a) [label=above:{1}] {\underline{\textbf{A}}};
\node[cascade] (casc) [right=of a] {Cascade}
edge[pre] (a);
\node[human] (bc) [right=of casc] [label=above:{2}] {B\underline{\textbf{C}}}
edge[pre] (casc)
edge[post, bend left] (a);
\node[invisible] [left=of a]{}
edge [slight] node[above, swap] {C} (a);

\end{tikzpicture}

\begin{Proof}
The only difference from the previous example is the inclination of the first player to
use the product A.
Due to that change, 
the first player will not refuse to use any of the products
when the incoming control edge is impossible to obey.
He will return to the use of the product A instead. 
This would entail the return of the cascade in the second state, 
the transition of the second player back to product C and infinite repetition of the cycle.

Thus, the constructed network is fragile.
\end{Proof}

Note that in the expanded network from the previous example, there exists a Nash equilibrium 
(total refusal of all players to choose any products), 
however, the only possible chain of individual improvements 
will not reach that equilibrium.

Also note that in this and the subsequent examples of endless cycles of individual improvements
(without global payoff requirements),
the main part of the cascade can be replaced by a single player of type 
A\textbf{ \underline{B}}. 
However, in this case we would also need external sources of inclination edges.

\subsection{An ineffective system}

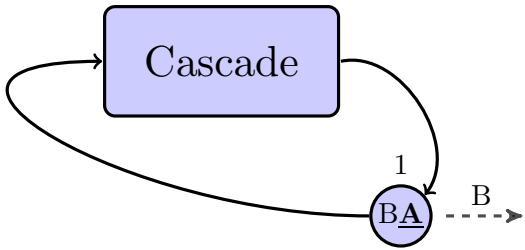
\begin{tikzpicture}
[bend angle=70]

\node[cascade] (casc) {Cascade};
\node[human] (ba) [ below right=10mm and 5mm of casc] [label=above:{1}] {B\underline{\textbf{A}}}
edge[pre, bend right] (casc.east)
edge[post, out=180, in=180, looseness=2 ] (casc.west);
\node[invisible] [right=13mm of ba]{}
edge [pslight] node[above, swap] {B} (ba);

\end{tikzpicture}

\begin{Proof}
In this example we need only one additional player, 
with inclination to A and under the emotional influence of the C-ranks.

Original equilibrium: the added player uses the product B.
Therefore, the cascade is in the first state, all the emotional edges are ``disabled''
(i.e. they push players to an impossible choice),
but the control edge holds the last player in the state B.
Thus, the network is indeed in an equilibrium.

Now forbid the added player to use the product B. 
Under the influence of the inclination edge, he will be forced to use the product A.
This switch will make the cascade transition into the second state.
All participating players will receive additional payoffs due to emotional edges.

Note that the resulting position is an equilibrium,
and would be an equilibrium in the original network. 
Indeed, after the cascade switches into the second state, 
the incoming control edge of the added player will push 
him to choose A and not B.
\end{Proof}

\subsection{An unsafe network}

\medskip

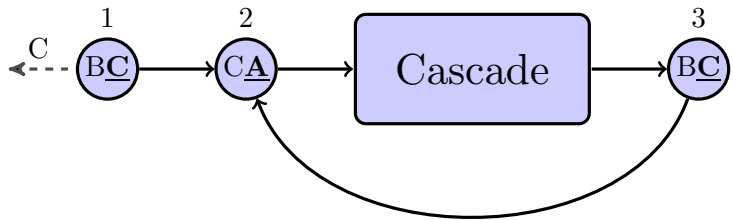
\begin{tikzpicture}
[bend angle=70]

\node[human] (bc1) [label=above:{1}] {B\underline{\textbf{C}}}
;
\node[human] (ca) [right=of bc1] [label=above:{2}] {C\underline{\textbf{A}}}
edge [pre] (bc1);

\node[cascade] (casc) [right=of ca] {Cascade}
edge [pre] (ca);
\node[human] (bc2) [ right= of casc] [label=above:{3}] {B\underline{\textbf{C}}}
edge [pre] (casc)
edge [post, bend left] (ca);
\node[invisible] [left=of bc1]{}
edge [pslight] node[above, swap] {C} (bc1);

\end{tikzpicture}

\begin{Proof}
This example largely repeats the fragile network. 
An additional player of type B\underline{C} prevents the infinite loop.
Obviously, forbidding this player to use the product C will destroy his effect on the system 
(because the next player has no access to the product B) and start the infinite chain of 
individual improvements.

It remains to verify that the initial network is in equilibrium. 
The first player has no incoming control edges, and therefore acts according to his 
inclination and selects the product C. 
The second player will be forced to choose product C, 
causing a cascade to switch into the first state, 
and the third player under will choose the product B. 
From this position, no individual improvement is possible,
and therefore it is indeed an equilibrium.

\end{Proof}

Now we can prove the following simple statement:

\begin{Theorem}
\emph{(a very bad network)}
There is a network that is vulnerable, fragile, inefficient and unsafe.
\end{Theorem}

\begin{Proof}
It suffices to combine the examples for vulnerable, fragile, inefficient and insecure network
into a single graph and distribute the emotional influence of the cascades from 
the first and the third examples to all external players 
(including internal players of other cascades).

Note that although in this case each player has incoming emotional edges
from multiple cascades, emotional invariant still holds, 
since no two cascades change their state simultaneously.
\end{Proof}

\section{``Paradoxes'' with the edge weight changes}
We now present some networks where analogous events
(global unidirectional payoff change,
infinite individual improvement chain) happen when 
changing the edge weights.

Such behaviour seems to be an interesting application of the
cascade construction, regardless of whether it is considered
paradoxical.

\subsection{Universal reduction of payoffs when reducing edge weight}

\medskip

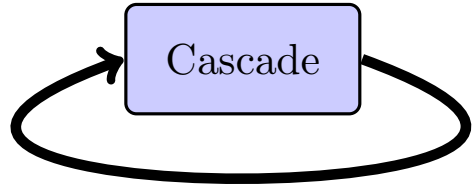
\begin{tikzpicture}

\node[cascade] (casc) {Cascade};
\draw [->, line width=4pt] (casc.east) to [out=-20, in =200, looseness=5 ] (casc.west);

\end{tikzpicture}

\medskip
\begin{Proof}
In this most simple example, it is enough to attach the outgoing 
control edge of the cascade as its own incoming control edge.
Note that in this case, the second state of the cascade is in equilibrium. 
If we reduce the weight of the loopback edge to zero, 
the cascade will return to the first state with the loss 
of emotional payoff for all the players. 
Note that the initial state of the cascade is an equilibrium 
for the original closed-loop network.
\end{Proof}

Note that the weight of the selected edges can be reduced 
to any value lower value $i-e$
(in this case, the incoming edge will never affect the strategic 
choice of a player even if it accidentally acts in the same direction as
emotional imbalance: the inclination will still be stronger).
So the same example can be used for both removal of an edge and 
for reduction of an edge's weight to some positive value.
The following examples will also use this fact.

\subsection{Universal payoff increase when reducing edge weight }

\medskip
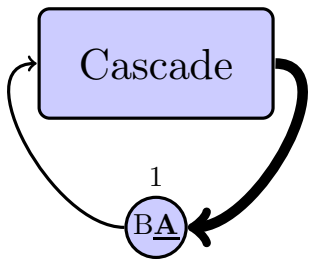
\begin{tikzpicture}
[bend angle=70]

\node[cascade] (casc) {Cascade};
\node[human] (ba) [ below =of casc] [label=above:{1}] {B\underline{\textbf{A}}}
edge[pre, line width=4pt, in=0, out=0] (casc.east)
edge[post, in=180, out=180] (casc.west);

\end{tikzpicture}

\medskip
\begin{Proof}
Here cascade is initially in the first state.
The cascade keeps an additional player in the state B,
which in turn allows 
the cascade to remain in the first state.

Now if we set the weight of the cascade's outgoing control edge
to zero, 
the additional player will be left without an incoming control edge 
and will go into the A state due to the inclination edge.
The cascade will switch into the second state, 
giving all participants an additional emotional payoff.
Note that this configuration is an equilibrium for the original network.
\end{Proof}

\subsection{Infinite loop when reducing edge weight}

\medskip
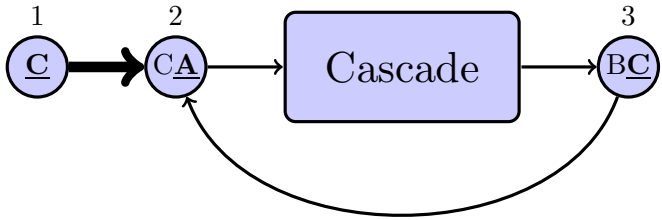
\begin{tikzpicture}
[bend angle=70]

\node[human] (c) [label=above:{1}] {\underline{\textbf{C}}}
;
\node[human] (ca) [right=of c] [label=above:{2}] {C\underline{\textbf{A}}}
edge [pre, line width=4pt] (c); 

\node[cascade] (casc) [right=of ca] {Cascade}
edge [pre] (ca); 
\node[human] (bc) [ right= of casc] [label=above:{3}] {B\underline{\textbf{C}}}
edge [pre] (casc)
edge [post, bend left] (ca);

\end{tikzpicture}

\medskip
\begin{Proof}
In this example we will use the same infinite loop design as for 
a fragile network. 
The equilibrium is maintained by a dedicated control edge
from the player using the product C. If this edge becomes zero-weight,
an infinite cycle of individual improvements will start.
\end{Proof}

\subsection{Universal payoff reduction when increasing edge weight}

\medskip

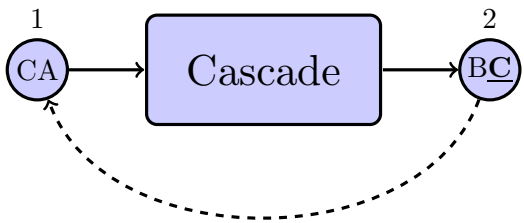
\begin{tikzpicture}
[bend angle=70]

\node[human] (ca) [label=above:{1}] {CA}
; 

\node[cascade] (casc) [right=of ca] {Cascade}
edge [pre] (ca); 
\node[human] (bc) [ right= of casc] [label=above:{2}] {B\underline{\textbf{C}}}
edge [pre] (casc)
edge [post, dashed, bend left] (ca);

\end{tikzpicture}

\medskip
\begin{Proof}
This example repeats the example of a vulnerable network, 
bu instead of allowing the product C to the first player, 
we add a control edge (marked on the diagram by a dotted line), 
making him switch to the product C, and remove the inclination 
for this product.
\end{Proof}

\subsection{Universal payoff increase when increasing edge weight}

\medskip

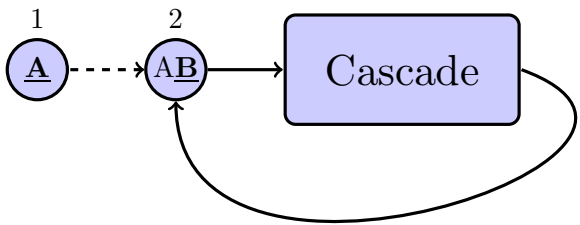
\begin{tikzpicture}
[bend angle=70]

\node[human] (a) [label=above:{1}] {\underline{\textbf{A}}}
; 
\node[human] (ab) [ right= of a] [label=above:{2}] {A\underline{\textbf{B}}}
edge [pre, dashed] (a); 
\node[cascade] (casc) [right=of ab] {Cascade}
edge [pre] (ab); 
\draw [->, very thick] (casc.east) to [out=-20, in =-90, looseness=1.8 ] (ab.south);

\end{tikzpicture}

\medskip
\begin{Proof}
In this example, the cascade is initially in its first state;
however, increasing the weight of the dotted connection from zero to $c$
forces the second player to move to state A.
The switch triggers the cascade transition into the second state.
Note that the resulting configuration is an equilibrium even in the original network.

\end{Proof}

\subsection{Infinite loop when increasing edge weight}

\medskip
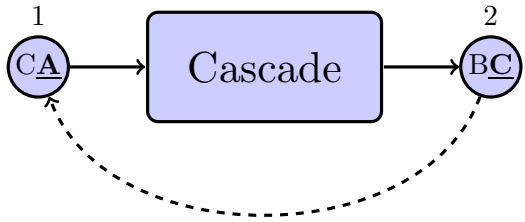
\begin{tikzpicture}
[bend angle=70]

\node[human] (ca) [label=above:{1}] {C\underline{\textbf{A}}}
; 

\node[cascade] (casc) [right=of ca] {Cascade}
edge [pre] (ca); 
\node[human] (bc) [ right= of casc] [label=above:{2}] {B\underline{\textbf{C}}}
edge [pre] (casc)
edge [post, dashed, bend left] (ca);

\end{tikzpicture}

\medskip
\begin{Proof}
In this example, we again use the same design of an infinite loop as previously.
We increase the weight of the dotted connection from zero to $c$ to
start the cycle.
\end{Proof}

\end{document}